\documentclass{article}
\usepackage{spconf,amsmath,graphicx,amsfonts}

\usepackage{lipsum}

\title{Analysis and Optimization of Aperture Design in Computational Imaging}
\name{Adam Yedidia, Christos Thrampoulidis, Gregory Wornell}
\address{
        Department of Electrical Engineering and Computer Science, MIT \\
        }
\usepackage[numbers]{natbib}

\usepackage{dsfont}

\usepackage{times}
\usepackage{hyperref}
\usepackage{url}
\usepackage{mathtools} 
\usepackage{cases}
\usepackage{subcaption}
\usepackage{amssymb,algorithm,cite,caption,cases,float,graphicx,url,color}
\usepackage{enumerate}
\usepackage{amsthm}

\definecolor{darkred}{RGB}{250,0,0}
\definecolor{darkgreen}{RGB}{0,150,0}
\definecolor{myblue}{RGB}{0,0,250}
\definecolor{darkblue}{RGB}{0,0,200}
\hypersetup{colorlinks=true, linkcolor=darkred, citecolor=myblue, urlcolor=darkblue}

\newcommand{\MI}{\mathcal{I}}
\newcommand{\MIo}{{\overline{\mathcal{I}}}}

\newcommand{\Q}{\mathbf{Q}}
\newcommand{\F}{\mathbf{F}}
\newcommand{\An}{\widetilde{\A}}

\newcommand{\Bern}[1]{\mathrm{Bern}(#1)}
\newcommand{\ds}{d^\star}

\newcommand{\MIr}{\MIo_{\sim1}}
\newcommand{\B}{\mathbf{B}}
\newcommand{\MIt}{\widetilde{\MIo}}

\usepackage{dsfont}

\newcommand{\simiid}{\stackrel{\text{iid}}{\sim}}

\newcommand{\Id}{\mathbf{I}}

\theoremstyle{theorem}
\newtheorem{propo}{Proposition}[section]

\newtheorem{lem}{Lemma}[section]

\theoremstyle{remark}
\newtheorem{remark}{Remark}

\theoremstyle{definition}

\DeclarePairedDelimiter\floor{\lfloor}{\rfloor}

\newcommand{\Exp}{\mathbb{E}}               
\newcommand{\E}{\mathbb{E}}                    
\newcommand{\la}{{\lambda}}                     

\newcommand{\nn}{\notag}

\newcommand{\A}{\mathbf{A}}

\newcommand{\ab}{\mathbf{a}}

\newcommand{\f}{\mathbf{f}}

\newcommand{\Nn}{\mathcal{N}}

\newcommand{\Oc}{\mathcal{O}}

\newcommand{\beq}{\begin{equation}}
\newcommand{\eeq}{\end{equation}}
\newcommand{\bea}{\begin{align}}
\newcommand{\eea}{\end{align}}

\newcommand{\vp}{\vspace{4pt}}

\newcommand{\D}{\mathbf{D}}

\begin{document}

%

\maketitle
\begin{abstract} 
There is growing interest in the use of coded aperture imaging systems
for a variety of applications.  Using an analysis framework
based on mutual information, we examine the fundamental limits of such
systems---and the associated optimum aperture coding---under simple
but meaningful propagation and sensor models.  Among other results, we
show that when thermal noise dominates, spectrally-flat masks, which
have 50\% transmissivity, are optimal, but that when shot noise
dominates, randomly generated masks with lower transmissivity offer
greater performance.  We also provide comparisons to classical pinhole cameras.

\end{abstract}
 


\section{Introduction}\label{sec:intro}

Digital signal processing plays an important role in modern imaging
systems. 
Many modern imaging systems operating
at optical and higher frequencies use coded apertures, whereby the
traditional lens in the aperture is replaced with a spatial mask that
selectively blocks portions of the light from reaching the sensor.
Yet while this is an increasingly important imaging modality---and one
with a long history dating back to the earliest pinhole
cameras---typical mask designs are guided by heuristics and/or
numerical procedures.  

\begin{figure}
\centering
\includegraphics[scale=0.15]{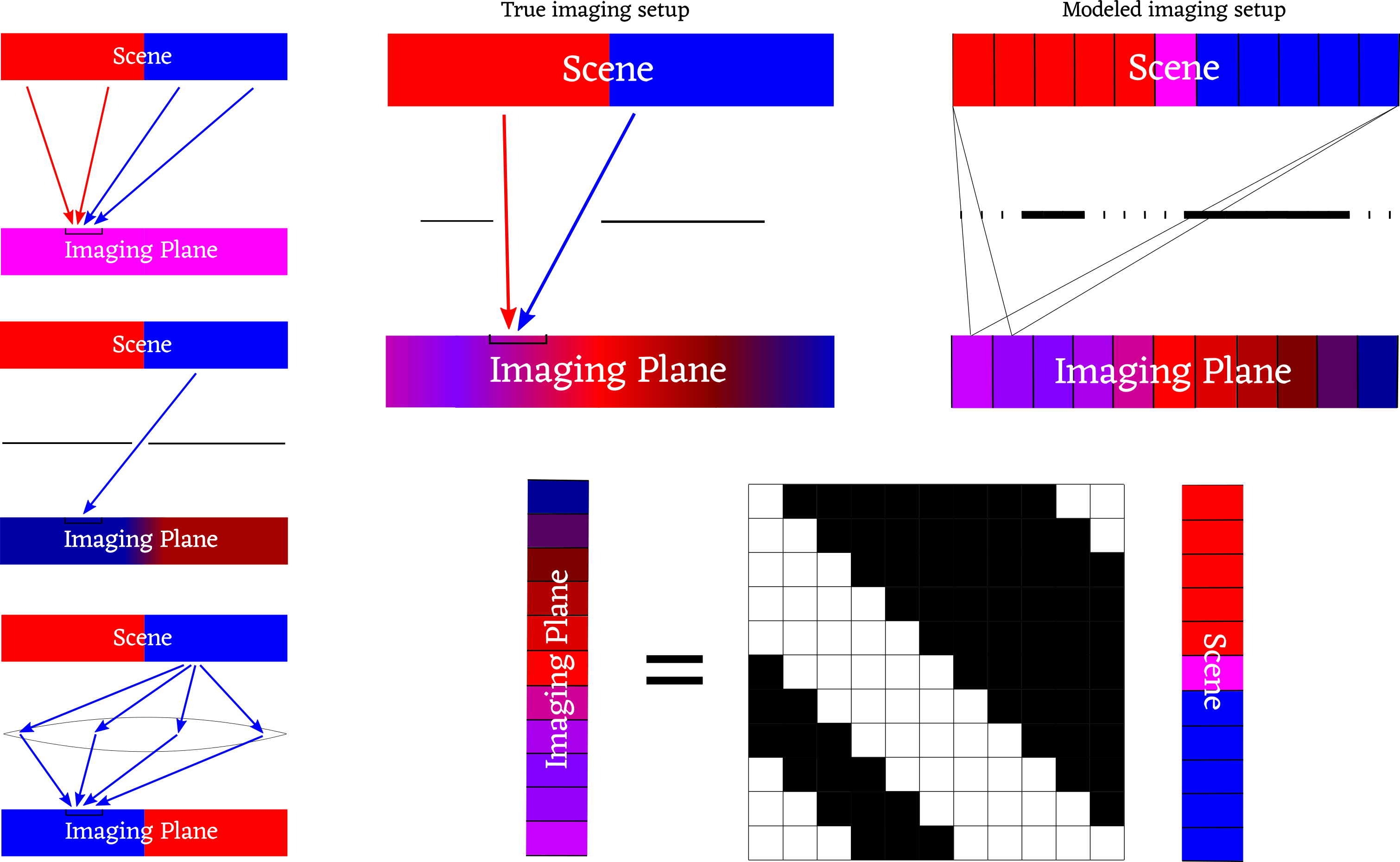}
\caption{Three imaging systems (left, top-to-bottom): no aperture,
  a pinhole and a lens.  Arrows indicate paths 
  light from the scene takes to a particular point on the imaging
  plane. On the right is an arbitrary mask, an illustration of its discretization
  and the corresponding transfer matrix.}
\label{fig:pinhole_illustration}
\end{figure}

As Figure \ref{fig:pinhole_illustration} depicts, with an empty
aperture, scene recovery from measurements at the imaging plane is
very poorly conditioned.  
Coded-aperture cameras seek to improve the conditioning of the problem through the use of more complicated (and transmissive) masks than a
pinhole in combination with suitably designed post-processing.

In this paper, we develop a comparative analysis of these imaging
systems, using mutual information as our performance measure.  Moreover, we use far-field geometric optics to model
propagation, and our sensor model at the imaging plane includes thermal and shot noise components.

Among the earliest and simplest instances of coded-aperture imaging
are those based on pinhole structure \citep{fenimore1978coded, young1971pinhole} and
pinspeck (anti-pinhole) structure \citep{cohen1982anti}, though more
complex structure is often used. Other methods involve cameras that uses a mask in addition to a lens to, e.g.,
facilitate depth estimation \citep{levin2007image}, deblur
out-of-focus elements in an image \citep{nayarcodedaperture}, enable
motion deblurring \citep{raskar2006coded}, and/or recover 4D
lightfields \citep{veeraraghavan2007dappled}.  Some forgo the lens altogether to decrease
costs and/or meet physical constraints \citep{duarte2008single}
\citep{asif2015flatcam}.  





Certain other systems, intended for non-line-of-sight applications,
rely on known structure in between the scene and the imaging
plane to improve the conditioning of the problem \citep{thrampoulidis2017exploiting}, like windows
\citep{windows} or corners of buildings \citep{wallcorners}.  These
can be viewed as instances of broader class of coded-aperture systems
that we analyze, in which the mask is naturally occurring and not chosen.

\section{Model}\label{sec:model}
\noindent{\textbf{Scene}}.~
Let $I(x)$ [W/m] represent the intensity of the scene over space in one dimension: $0\leq x\leq L$. We denote $J=\int I(x)\mathrm{d}x$ [W] the net power radiated. Assume a uniform discretization of $[0,L]$ into $n$ bins of size $\Delta=L/n$ each, and denote $x_1,x_2,\ldots,x_n$ their centers. We assume that the discretization is fine enough that the intensity at each bin $i\in[n]$ takes constant value $I(x_i)$. Let $f_i=I(x_i)\cdot \Delta$ be the power radiated from each bin.
We model $\f=[f_1,\ldots,f_n]$ as a multivariate Gaussian distribution $\Nn({\mu}\mathbf{1},\Q)$ with mean $\mu$ and covariance matrix $\Q$. We set $\mu=J/n$ to ensure that the average net power is $E[\sum_{i\in[n]}f_i]=\sum_{i\in[n]}\mu=J$. The Gaussian statistics model for images is frequently used, such as in \citep{gaussianmodel,levin2007image}. 
In this paper, we consider the following two cases:
%

\noindent{\emph{IID:}}~
We assume that the $f_i$'s are uncorrelated, i.e., $\Q=\mathbf{I}$. While natural scenes will exhibit correlations, studying the IID case is a means of performing a worst-case analysis.

\noindent{\emph{$1/f$-prior:}}~
We follow a classical statistical model according to which the power spectrum of natural images depends as $1/f$ over the spatial frequency \citep{frequencyprior}, by taking $Q = \F_n^*\D^\star\F_n$, where $\F_n$ is the normalized DFT matrix of size $n$ and $\D^\star$ is a diagonal matrix with the following entries:
$\ds_i=\ds_{n/2+i}=1/i,$ for $i=1,\ldots \floor{n/2}.$

\noindent{\textbf{Imaging plane}}.~  The imaging plane consists of $m$ adjacent and equally-sized pixels. We focus on the case where $m=n$. The power $y_j$ [W] measured at each pixel is $y_j=\frac{1}{n} \sum_{i=1}^n A_{ji}\cdot f_i$, where $f_i$ is the power radiated from the $i^\text{th}$ bin. 
The $(1/m)$--scaling is chosen to ensure preservation of energy:
$\E[\sum_j{y_j}] = \frac{1}{m}\sum_j\sum_i A_{ji} \cdot \E[x_i]  \leq \frac{1}{m}\cdot m n \cdot \frac{J}{n} = J.$
The measurement model is a reduction of a more complete forward model, which further accounts for distance attenuation and cosine factors in light propagation \citep{brdf}.
This reduction corresponds to a scenario in which the scene is far enough from the imaging plane that the distance attenuation and cosine factors are well-approximated by constants. 
%

\noindent{\textbf{Aperture}}.~Denote by $\A$ the $m\times n$ transfer matrix whose entries $\A_{ji}$ model the aperture. We assume that a maximal integration time is allowed, and normalize it so that the maximal value for each entry of $\A$ is $1$. We let $\rho$ denote the \emph{transmissivity} of the aperture. For an on-off aperture, $\rho$ measures the fraction of elements that transmit light (See Fig.~\ref{fig:pinhole_illustration}). In general, we assume a circulant $\A$; that is equivalent to assuming that the mask repeats a certain pattern (of length $n$) twice: 
 $
 A_{ji} = a_{(i-j)\mod{n}}
 $
 where $\ab^T = (a_0,\ldots,a_{n-1})^T$ is the first row of $\A$.  


\noindent{\textbf{Noise}}.~ We distinguish between two different types of noise.

\noindent{\emph{(Thermal noise):}}~
This includes noise sources that are independent of the contribution to the measurements due to the scene of interest. We model it as additive Gaussian with variance $W/m$, i.e., constant net noise power $W$ and each pixel absorbs power proportional to its size, giving rise to the $1/m$ factor.

\noindent{\emph{(Shot noise):}}~
This includes measurement noise that depends on the contribution due to the scene of interest. This results in additive Gaussian noise of variance  $\rho\cdot\frac{J}{m}$ (proportional to the net power of light that goes through the aperture). 

\noindent Overall, the measurement at each pixel is modeled as 
$y_j = ({1}/{m})\sum_{i\in[n]}A_{ji}  f_i + z_j,$
where  $z_j\sim\Nn(0,(W+\rho\cdot J)/m)$.




\noindent{\textbf{Mutual information}}.~ The  mutual information (MI) between the measurements $y_j, j\in[m]$ and the unknowns $f_i, i\in[n]$ of the imaging problem is given as 
$\MI = 
\log\det\big( \frac{1}{W+\rho\cdot J}\cdot\frac{1}{m}\cdot\A\Q\A^T + \Id \big).$
  Recall that a circulant matrix is diagonalized by $\F_n$. Also,  $\Q=\F_n^*\D\F_n$ where $\D=\mathbf{I}$ (IID scene) or $\D=\D^\star$ (1/f-prior). With these, $\MI$ reduces to (recall  $m=n$)
\begin{align}\label{eq:MI_def2}
\MI = 
\sum_{i=1}^{n}\log\big( \frac{1}{W+\rho\cdot J}\cdot{d_{i}}\cdot \frac{|\la_i(\A)|^2}{n} + 1 \big),
\end{align}
where, $\la_i(\A)$ denotes the eigenvalue of $\A$ corresponding to the $i^\text{th}$ frequency. We often write $\la_i$ when clear from context. 

\noindent{\textbf{Aperture Types}.}~ Here, we summarize several types of aperture designs and their corresponding models.  

\noindent{\emph{Pinhole:}}~ We model a pinhole camera as an on-off mask with only a single open element, i.e., $\A=\mathbf{I}$ (or, any permutation of the identity). Also, for a pinhole: $\rho=1/n$. 

\noindent{\emph{Spectrally-Flat patterns:}}~ The family includes pseudo-noise binary (0/1) patterns such as maximum length sequences (MLS) and uniform redundant array patterns such as URA and MURA. Onwards, we refer to patterns with the following properties as spectrally-flat patterns:  (i) $\rho\approx 1/2$ (there is one more one than zero);  (ii) they are spectrally flat with the exception of a DC term \citep{MLS,MURA,review,busboom1998uniformly}.

\noindent{\emph{Random on-off patterns:}}~We study random patterns where each entry of $\ab$ is generated IID $\Bern{p}$, for $p\in(0,1]$. 
For such random on-off patterns we use $\rho=p$, since for large $n$ (which is our focus) the number of on-elements is $\approx np$.

\noindent{\emph{Random uniform patterns:}}~ We also study patterns consisting of elements that can partially absorb light, e.g., \citep{glass, veeraraghavan2007dappled}. We focus on random such patterns where each entry of $\ab$ is IID Uniform$([0,1])$. For these patterns, the expected transmissivity $\rho=1/2$.    


\section{Results}
\subsection{IID scene}
Throughout this section we study the IID scene model. It is convenient to work with the \emph{normalized} mutual information per pixel $\MIo:=\MI/n.$

\vspace{-10pt}
\subsubsection{Pinhole}
\vspace{-5pt}
From \eqref{eq:MI_def2} the (normalized) MI of a pinhole is given by
$
\MIo_{\text{pinhole}} = \log\big( \frac{1}{n\cdot W+J} + 1 \big).
$
By allowing only a fraction of $1/n$ of the light to go through, the formula justifies that the performance of a pinhole deteriorates drastically for large $n$ (cf., MI goes to zero, unless $W$ becomes negligible, e.g., unless it scales inversely proportionally to $1/n$). 
Note that this result applies only to a vanishingly small pinhole (decreasing in size as $n$ increases); a pinhole of fixed size achieves constant mutual information per pixel. 



\subsubsection{Spectrally-flat patterns}
The following proposition characterizes the MI of spectrally-flat patterns and shows that they maximize MI when thermal noise is dominant. See Appendix \ref{sec:proof} for a proof sketch.

\begin{propo}\label{lem:SF} Consider the IID scene model. Let $\MI_{\star}$ be the mutual information of a spectrally-flat pattern for an odd $n$.\footnote{Here, we implicitly assume that $n$ is such that an MLS, or URA, or MURA pattern exists. For example, MURA patterns can be generated for any prime $n$ that is of the form $4d+1,~d=1,2,\ldots$.} It holds that:
\begin{align}\label{eq:IID_spec_flat}
\lim_{n\rightarrow\infty}\MIo_{\star}=
 \log\big(\frac{1/4}{W+J/2}+1\big).
\end{align}
Moreover, if $W\gg J$, then given the mutual information $\MIo_p$ of any on-off aperture design with $np$ ``on" elements and $p\neq \frac{1}{2}$, for large enough $n$, it holds that 
$\MIo_p< \MIo_{\star}.$
\end{propo}

\begin{remark}
For spectrally-flat occluders, $\la_1\approx \frac{n}{2}$ and $|\la_2|=\ldots=|\la_n|\approx\frac{\sqrt{n}}{2}.$ The contribution of the first eigenvalue to the sum in \eqref{eq:MI_def2} is $\Oc(\log(n)/n)$, which captures the rate at which convergence in \eqref{eq:IID_spec_flat} is true. Throughout, statements that involve $n\rightarrow\infty$ are to be interpreted with the rest of parameters (such as $W$, $J$, $\rho$) held constant (independent of $n$).
\end{remark}

\begin{remark}
The advantage of spectrally-flat apertures when thermal noise is dominant is shown by concavity of $\log$ and applying Jensen's inequality to \eqref{eq:MI_def2}. 
More generally, Jensen's inequality is tight iff $\la_2 d_2=\la_3 d_3=\ldots=\la_n d_n$. This leads to optimality of flat-spectrum patterns for $d_i=1$ (IID scenes), but the conclusion might be different for correlated scenes. Also, we require that $W\gg J$. In the next section, we show that if this is not the case then certain patterns with $p<1/2$ can outperform the spectrally flat ones.
\end{remark}

\vspace{-10pt}
\subsubsection{Random on-off patterns}\label{sec:on/off}
%
\vspace{-5pt}
We explicitly compute the asymptotic value of the MI for random on-off patterns.  Our theoretical results use tools from random matrix theory (RMT) \citep{RMT_circ1,RMT_circ2} and are thus asymptotic in nature. (However, numerical simulations suggest accuracy of the predictions for $n$ on the order of a few hundreds.) 
A proof sketch is deferred to Appendix \ref{sec:proof}.

\begin{propo}
\label{lem:IID_rand}
 Assume the IID scene model. Let 
$X$ be a random variable with density $f_X(x)=|x|e^{-x^2}$. The mutual information $\MI_p$ for a random on-off circulant system with parameter $0<p<1$ converges in probability with $n$ to:
$\MIt_p = \E_X[\log(\frac{p(1-p)}{W+pJ} X^2 + 1)].$
\end{propo}

\begin{figure}
\centering
\includegraphics[scale=0.4]{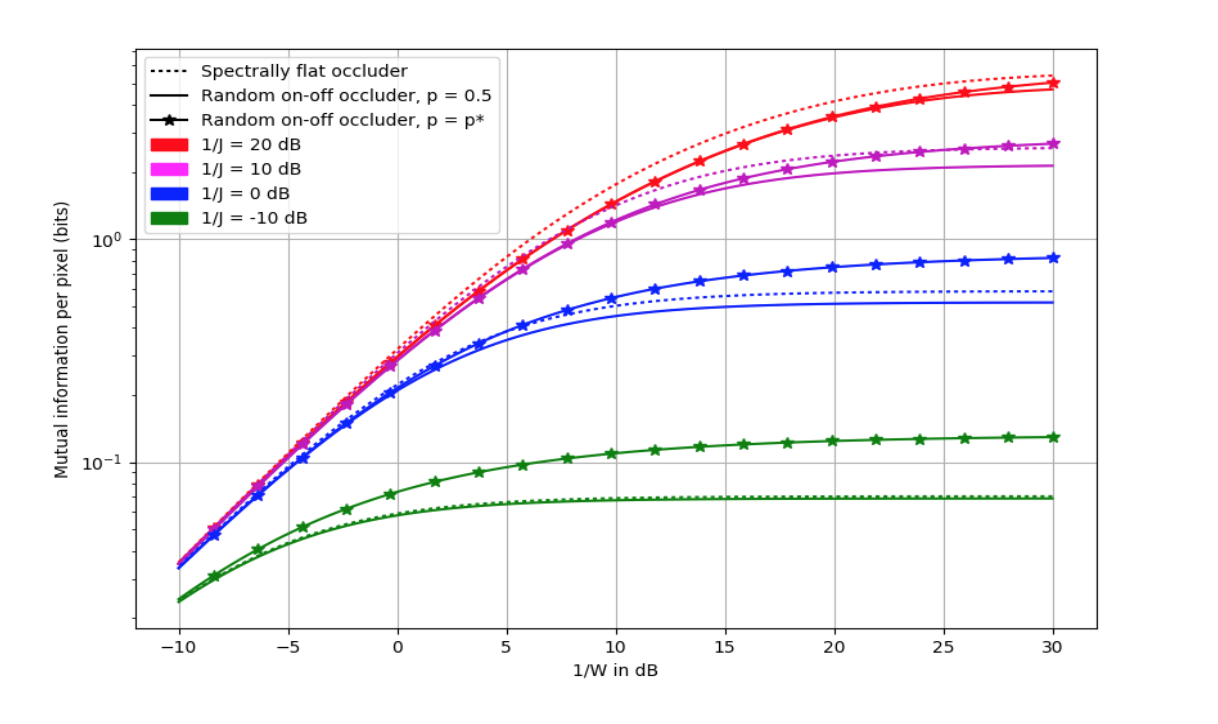}
\caption{A plot of the MI per pixel of a spectrally-flat occluder, a random on-off occluder with $p=0.5$, and a random on-off occluder with optimally chosen $p^\star$. See Proposition \ref{lem:IID_rand}.}
\label{fig:pinhole}
\end{figure}

\begin{remark}\label{rem:p}
Maximizing the formula of the proposition over $p$ gives the optimal choice of the transmissivity parameter. Since $\log$ is increasing, it can be shown that the maximum occurs at 
\begin{align}\label{eq:p_star}
p^\star = ({W}/{J})\cdot( \sqrt{1+J/W} - 1 ).
\end{align}
In particular, when ambient noise is dominant ($W\gg J$), then using $\sqrt{1 + \frac{J}{W}} \approx 1 + \frac{J}{2W}$ gives $p_\star\approx\frac{1}{2}$. On the other hand, when  shot noise is dominant ($J\gg W$), then $p_\star\approx\sqrt{\frac{1}{J}}$; thus, fewer open holes in the aperture design are desirable. See Figure \ref{fig:pinhole} for an illustration. For small values of $1/W$ (relative to $1/J$): $\MIt_{p^\star}\approx \MIt_{\frac{1}{2}}$, but $\MIt_{p^\star}> \MIt_{\frac{1}{2}}$ when $1/J$ is small.
\end{remark}

\begin{remark}
For $\Bern{1/2}$ patterns, an application of Jensen's inequality verifies that
$
\MIt_{\frac{1}{2}} < \log(\frac{1/4}{W+J/2} \E_X[X^2] + 1) = \MIo_\star,
$
i.e., spectrally-flat patterns are superior. On the other hand, a random pattern $\Bern{p^\star}$ with optimal parameter given by \eqref{eq:p_star} can outperform the spectrally-flat one. For example, this happens when shot noise is dominant, as illustrated in Figure \ref{fig:pinhole}. In the same figure, spectrally-flat patterns are superior when $W\gg J$ as predicted by Proposition \ref{lem:SF}.
\end{remark}

\subsubsection{Random uniform patterns}
Similar to Proposition \ref{lem:IID_rand} we leverage results of \citep{RMT_circ1} to evaluate the MI performance of random uniform patterns; we omit the details due to space limitations.
\begin{propo}
\label{lem:IID_unif}
 Consider the IID scene model. Let 
$X$ be a random variable with density $f_X(x)=|x|e^{-x^2}$. The normalized mutual information $\MIo_\text{uniform}$ for a random uniform circulant system  converges in probability with $n$ to: 
$\MIt_\text{uniform} = \E_X[\log(\frac{1/24}{W+J/2} X^2 + 1)].$
\end{propo}

Comparing the formula of the proposition to Proposition \ref{lem:IID_rand}, reveals that
$\MIt_\text{uniform}<\MIt_p, \quad\text{for all}\quad p\in[\frac{1}{2}-\frac{1}{\sqrt{6}},\frac{1}{2}].$
Hence, random on-off masks in this range of $p$ outperform random uniform masks. In short, if physical limitations prevent the use of apertures that can redirect light, but can only absorb it, then absorbing all (with appropriate $p$) is better than partially (at least for random designs).



\subsection{Correlated scene}\label{sec:corr}
 
We extend the ``worst-case" analysis of the previous section regarding IID scenes to correlated ones. We follow the 1/$f$ scene prior model. Due to space limitations, we restrict the exposition to spectrally-flat  and random on-off patterns.

\vp
\noindent{\emph{Spectrally-flat patterns:}}~
 The MI of the spectrally-flat patterns for correlated scenes can be computed similar to \eqref{eq:IID_spec_flat}. For large enough $n$, we find that 
$\MI^\star\approx\log(\frac{n/4}{W+J/2}+1) + 2\sum_{k=2}^{\frac{n-1}{2}}\log(\frac{1/4}{W+J/2}\frac{1}{k}+1)
\approx \log(\frac{1/4}{W+J/2} \cdot n) + \frac{1/2}{W+J/2}(\log(n/2)-1),$
where, for the first approximation: $\frac{n-1}{n}\approx 1$, and, for the second one: $\log(1+x)\approx x$ for $|x|\ll 1$ and $\sum_{k=1}^n{\frac{1}{n}}\approx\log{n}$.
In contrast to the IID case where the MI scaled linearly with $n$, here it scales as $\mathcal{O}(\log(n))$.

%
\vp
\noindent{\emph{Random on-off patterns:}}~
Contrary to the case of IID scenes where knowledge of the the limiting spectral density of $\A$ suffices to characterize the MI, for correlated scenes each eigenvalue is weighted differently. Hence, the behavior of the MI depends on the statistics of each individual eigenvalue. Since $\A$ is circulant, the eigenvalues are exactly the Fourier coefficients of the entries  of the generating vector $\ab$, i.e., $\la_1=\sum_{\ell=0}^{n-1}a_\ell$, and, for $k=2,\ldots,\frac{n-1}{2}$ (assume $n$ is odd for simplicity): $\la_k^2 = \la_{n-k}^2 = g_k^2 + h_k^2$, where
$g_k :=  \sum_{\ell=0}^{n-1}a_\ell\cdot \cos(\ell k \frac{2\pi}{n}),
\quad h_k :=  \sum_{\ell=0}^{n-1}a_\ell\cdot \sin(\ell k \frac{2\pi}{n}) .$
Next, observe that if the $a_i$'s were standard Gaussians then the following statements hold. (a) $\lambda_1$ is distributed $\Nn(0,{n})$. (b) $g_k$'s and $h_k$'s are IID $\Nn(0,1/2)$; therefore, $\la_k^2\simiid\frac{1}{2}\chi_2^2$ where $\chi_2^2$ denotes a chi-squared random variable with two degrees of freedom. This leads to the following conclusion:

\begin{lem}\label{lem:Gauss}
Let the first row of a circulant $\A$ have entries drawn IID standard Gaussians and the MI be given as in \eqref{eq:MI_def2}, for some $\gamma:=\frac{1}{W+\rho\cdot J}$ and for $d_i=d_i^\star$. Then, $\E[\MI]$ equals
$\Exp_{G\sim\Nn(0,1)}\log\big( \gamma n G^2 +1\big) +2 \sum_{k=2}^{\frac{n-1}{2}} \Exp_{X\sim\chi_2^2}\log\big( \gamma \frac{X}{2i} + 1 \big).$

\end{lem}

\begin{figure}
\centering
\includegraphics[scale=0.45]{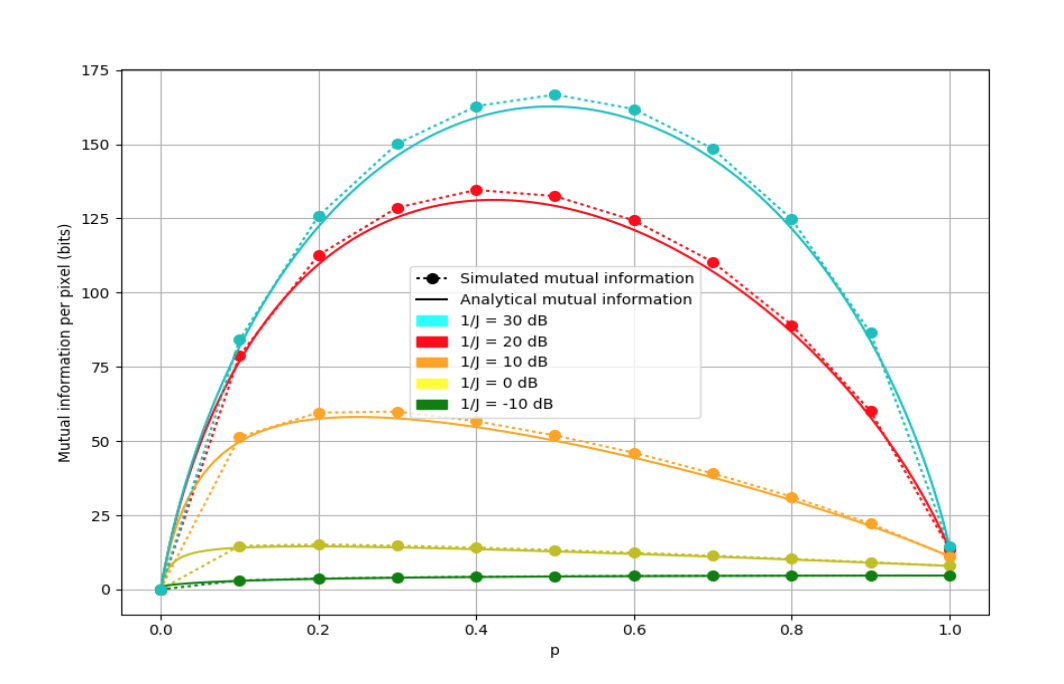}
\caption{Analytical formula follows Eqn.~\eqref{eq:exp_Bern}. Simulated data are averages of 1000 randomly generated apertures of size  $n=250$ for various different values $p$. We set $W = -20 dB$.}
\label{fig:correl_comp}
\end{figure}

We conjecture that the conclusion of Lemma \ref{lem:Gauss} is universal over the distribution of the entries of $\ab^T$, i.e., it holds for entries that have zero mean, unit variance, and bounded third moment. Based on this assumption, we conjecture that the expected mutual information $\E[\MI_p]$ for a random on-off circulant system with parameter $0<p<1$ for the correlated scene model is given by:
{\small
\begin{align}%
\nn
&\Exp_{G\sim\Nn(0,1)}\log\big( \frac{(\sqrt{p(1-p)}\cdot G + p\sqrt{n})^2}{W+pJ} +1\big) \\
&\qquad\qquad+ 2 \sum_{k=2}^{\frac{n-1}{2}} \Exp_{X\sim\chi_2^2}\log\big( \frac{p(1-p) X}{W+pJ}\frac{1}{2k} + 1 \big).    \label{eq:exp_Bern}
\end{align}
}
Figure~\ref{fig:correl_comp} shows a comparison of the formula predicted by \eqref{eq:exp_Bern} against simulated data. It further reveals that \eqref{eq:exp_Bern} can be used to numerically evaluate the optimal $p = p^*$.

\section{Discussion and Future Work}


Our framework allows to rigorously show that spectrally-flat patterns are optimal for IID scenes, and formalize the arguably unintuitive empirical claim (discussed, for instance, in~\citep{levin2007image}) that the best masks tend to transmit half the light they receive. \citep{veeraraghavan2007dappled} raises the question of whether continuous-valued masks perform better than binary-valued ones; we plan to use our framework to find an answer in the future.
In this work, we focused exclusively on 1D masks, which are relevant for example in de-blurring along one dimension~\citep{raskar2006coded}. We leave extensions to 2D masks to future work. However, we mention in passing that that much of the analysis conducted here can be directly applied to study separable 2D apertures, i.e. ones that can be expressed as the outer product of two 1D apertures.

\appendix


\section{Proof sketches}\label{sec:proof}

\noindent{\emph{Proof sketch of Proposition \ref{lem:SF}}:}~
For convenience set $\la_i:=\la_i(\A)$ and $\gamma=\frac{1}{W+pJ}$. We treat the DC-term of the spectrum, i.e. $\lambda_1$, separately from the rest. Note that $\A\mathbf{1} = (np)\mathbf{1}$;s thus, $\la_1=np.$ 
Next, let us denote $\MIr$ the MI in \eqref{eq:MI_def2}, excluding the term that involves $\la_1$. By concavity of $\log$ and Jensen's inequality, $\MIr$ is upper bounded by 
{\small
\begin{align}
\frac{n-1}{n}\log\big( \frac{\gamma}{(n-1)n}\sum_{i=2}^n {|\la_i(\A)|^2} + 1 \big)\approx \log\big( \gamma p(1-p) + 1 \big), \label{eq:bulk_det}
\end{align}
}
where the bound is tight iff $|\la_2|=|\la_3|=\ldots=|\la_n|$; \eqref{eq:bulk_det} uses the fact that $\sum_{i=2}^n {|\la_i(\A)|^2}=\|\A\|_F^2-\la_1^2=n^2p(1-p)$ and $n\approx n-1$ for large $n$. In particular, spectrally-flat patterns achieve the upper bound, which gives
$
\MIo_\star \approx \frac{1}{n}\log(\gamma\frac{n}{4} + 1) + \log\big(\frac{\gamma}{4} + 1 \big) \stackrel{n\rightarrow\infty}{\longrightarrow} \log\big(\frac{\gamma}{4} + 1 \big).
$
Next assume $W\gg J$ such that $\gamma\approx\frac{1}{W}$. The upper bound in \eqref{eq:bulk_det} is then  maximized for $p=1/2$. On the other hand, the contribution of $\la_1$ is at most $\frac{1}{n}\log(\gamma n^2+1)$, which goes to zero for large $n$. 

\noindent{\emph{Proof sketch of Proposition \ref{lem:IID_rand}:}~
The proof leverages the following result of \citep{RMT_circ1}. Consider a reverse circulant matrix $\frac{1}{\sqrt{n}}\B$ with entries $B_{ji}=b_{j+i-2\mod n}$ and $(b_0,b_1,\ldots,b_n)$ a sequence of IID random variables with mean zero, unit variance and bounded third moment. Then, the empirical spectral density (ESD) of $\B$ converges to the limiting spectral distribution with density $f_X(x)$. 
In our setting, we are interested on the ESD of $\A\A^T$ for $\A$ that has entries $\Bern{p}$. To apply the result of \citep{RMT_circ1}, consider:
$
\An=(\A-p\mathbf{1}\mathbf{1}^T)/\sqrt{p(1-p)}.
$
The entries of $\An$ have now zero mean and unit variance. Moreover, $\la_j(\An)=\la_j(\A)/\sqrt{p(1-p)}$ for $j=2,\ldots,n$. It can be shown  that $|\la_j(\An)|^2=\la_j^2(\B)$ \citep[Lem.~1]{RMT_circ1}. Applying these to \eqref{eq:MI_def2} gives $\MI = \frac{1}{n}\sum_{i=1}^{n}\log\left( \frac{p(1-p)}{W+pJ}\cdot \la^2_i\big(\frac{1}{\sqrt{n}}\B\big) + 1 \right)\stackrel{n\rightarrow\infty}{\rightarrow} \MIt_p,$
where the convergence result follows from \citep{RMT_circ1}.

%

%


\bibliographystyle{IEEETran}
\bibliography{REVEAL}

%
%

\end{document}